\documentclass[11pt]{article}
\usepackage[utf8]{inputenc}
\usepackage[T1]{fontenc}
\usepackage{lmodern}
\usepackage{amsmath,amssymb}
\usepackage{booktabs}
\usepackage{hyperref}
\usepackage[margin=1in]{geometry}

\title{Zero-Cost NDV Estimation from Columnar File Metadata}
\author{Claude Brisson\\\texttt{claude.brisson@gmail.com}}
\date{}

\begin{document}
\maketitle

\begin{abstract}
We present a method for estimating the number of distinct values (NDV) of a column in columnar file formats, using only existing file metadata--no extra storage, no data access. Two complementary signals are exploited: (1)~inverting the dictionary-encoded storage size equation yields accurate NDV estimates when distinct values are well-spread across row groups; (2)~counting distinct min/max values across row groups and inverting a coupon collector model provides robust estimates for sorted or partitioned data. A lightweight distribution detector routes between the two estimators. While demonstrated on Apache Parquet, the technique generalizes to any format with dictionary encoding and partition-level statistics, such as ORC and F3. Applications include cost-based query optimization, GPU memory allocation, and data profiling.
\end{abstract}

\section{Introduction}

While building Theseus~\cite{theseus}, a GPU-accelerated distributed query engine at VoltronData, we needed NDV estimates to drive cost-based optimization--aggregate pushdown, join ordering, memory allocation for GPU kernels. Parquet's \texttt{distinct\_count} field is almost never populated: computing exact distinct counts is expensive, and most writers omit it. Sampling or maintaining HyperLogLog sketches would require accessing data or adding writer-side infrastructure, defeating the purpose of metadata-only planning.

The question became: \emph{what cardinality information is already encoded in existing metadata?}

Dictionary-encoded columns store their uncompressed size per column chunk. Row groups store per-column min/max statistics. Neither was designed for cardinality estimation, but both encode it implicitly--at least locally, within individual row groups. This paper details techniques for generalizing these local signals into global NDV estimates for the entire column.

The technique was implemented and validated on production workloads at VoltronData. When the company's assets were liquidated, the implementation and experimental results were lost; this paper reconstructs the approach from memory.

\paragraph{Contributions.}
\begin{enumerate}
\item A closed-form equation relating NDV to dictionary-encoded storage size, solved via Newton-Raphson (Section~\ref{sec:dict}).
\item Recognition that row group min/max statistics function as implicit cardinality sketches, with NDV recovery via coupon collector inversion (Section~\ref{sec:minmax}).
\item A lightweight distribution detector that routes between the two estimators (Section~\ref{sec:distribution}).
\item Batch memory prediction using coupon collector analysis (Section~\ref{sec:batch}).
\end{enumerate}

\section{Background}
\label{sec:background}

\subsection{Parquet Dictionary Encoding}

Apache Parquet organizes data into \emph{row groups}, each containing one \emph{column chunk} per column. Within a column chunk, Parquet employs dictionary encoding for low-to-medium cardinality columns~\cite{parquet}: a dictionary page stores the distinct values, and data pages store bit-packed indices into the dictionary.

\begin{verbatim}
Dictionary Page          Data Pages
+-------------+          +---+---+---+---+---+
| "alice"   0 |          | 0 | 2 | 1 | 0 | 2 | ...
| "bob"     1 |          +---+---+---+---+---+
| "charlie" 2 |          (indices, bit-packed)
+-------------+
\end{verbatim}

Each column chunk exposes a \texttt{total\_uncompressed\_size} field in its metadata, equal to the sum of dictionary and index pages before compression:
\begin{equation}
S = \text{ndv} \times \text{len} + (N - \text{nulls}) \times \lceil \log_2(\text{ndv}) \rceil \,/\, 8
\label{eq:storage}
\end{equation}
where $\text{ndv}$ is the number of distinct values, $\text{len}$ the mean byte length of values, $N$ the row count, and $\text{nulls}$ the null count.

\subsection{Row Group Statistics}

Parquet stores per-column min/max statistics in each row group~\cite{parquet-stats}. These are designed for predicate pushdown--skipping row groups whose value ranges exclude a query predicate--but have not traditionally been used for cardinality estimation.

Across $n$ row groups, a column yields $n$ pairs $(\min_i, \max_i)$. We exploit both the diversity of these values and the overlap patterns between consecutive ranges.

\section{Approach Overview}
\label{sec:overview}

The two metadata signals have complementary strengths. Dictionary size inversion (Section~\ref{sec:dict}) is grounded in a storage equation whose accuracy depends on how distinct values are spread across row groups. It does not require statistical uniformity--values may have arbitrary frequency distributions within each row group. What matters is that most distinct values appear in most row groups, a condition we call \emph{well-spread}. When row group ranges heavily overlap, this condition holds and dictionary inversion is accurate. Min/max diversity estimation (Section~\ref{sec:minmax}) instead exploits the spread of extreme values across row groups; it is accurate for sorted or partitioned data, where dictionary inversion underestimates.

\begin{table}[h]
\centering
\begin{tabular}{lccc}
\toprule
& Well-spread & Sorted/Partitioned & Low NDV \\
\midrule
Dictionary inversion & accurate & underestimates & accurate \\
Min/max diversity & underestimates & accurate & underestimates \\
\bottomrule
\end{tabular}
\caption{Complementary accuracy profiles of the two estimators.}
\label{tab:profiles}
\end{table}

A distribution detector (Section~\ref{sec:distribution}) classifies each column's data layout by analyzing range overlap and monotonicity across row groups. The final estimate takes the maximum of both methods, bounded by type-specific constraints (Section~\ref{sec:combining}).

\section{Dictionary Size Inversion}
\label{sec:dict}

\subsection{Core Equation}

Given the observable metadata tuple $(S, N, \text{nulls}, \text{len})$, we solve Equation~\eqref{eq:storage} for $\text{ndv}$:
\begin{equation}
f(\text{ndv}) = \text{ndv} \times \text{len} + (N - \text{nulls}) \times \lceil\log_2(\text{ndv})\rceil \,/\, 8 - S = 0
\label{eq:root}
\end{equation}

\subsection{Newton-Raphson Solution}

We apply Newton-Raphson iteration using the exact $f$ but a continuous approximation for the derivative, since the ceiling function has zero derivative almost everywhere:
\begin{equation}
\text{ndv}_{k+1} = \text{ndv}_k - \frac{f(\text{ndv}_k)}{f'(\text{ndv}_k)}, \qquad f'(\text{ndv}) \approx \text{len} + \frac{N - \text{nulls}}{8 \cdot \text{ndv} \cdot \ln 2}
\end{equation}

The ceiling discontinuities are small relative to the dominant linear term, and convergence is unaffected in practice. The initial guess $\text{ndv}_0 = S / \text{len}$ assumes index overhead is small. Convergence to a tolerance of $10^{-6}$ is typically achieved in 5--10 iterations.

\subsection{Estimating Mean Value Length}
\label{sec:len}

The parameter $\text{len}$ is not directly stored in metadata. For fixed-width types (integers, dates, timestamps), it is known exactly from the Parquet schema. For variable-length types, we estimate it from all distinct min and max values observed across row groups:
\begin{equation}
\text{len} = \frac{1}{|\mathcal{V}|} \sum_{v \in \mathcal{V}} |v|, \qquad \mathcal{V} = \{\text{distinct mins}\} \cup \{\text{distinct maxs}\}
\end{equation}

The sample size $|\mathcal{V}|$ serves as a reliability indicator: more distinct extreme values yield a more representative length distribution. For a single row group, we fall back to $\text{len} = (|\min| + |\max|) / 2$.

\subsection{Plain Encoding Fallback}
\label{sec:fallback}

When a column's cardinality exceeds Parquet's dictionary threshold (typically ${\sim}1$\,MB), the writer falls back to plain encoding~\cite{parquet}: no dictionary page is written, and data pages contain raw values instead of indices. The uncompressed size then satisfies $S \approx (N - \text{nulls}) \times \text{len}$, which the solver interprets as very high NDV.

We detect this condition by checking two indicators simultaneously:
\begin{equation}
\text{likely\_fallback} = \left(\frac{\text{ndv}}{N - \text{nulls}} \geq 0.9\right) \;\land\; \left(\frac{S}{(N - \text{nulls}) \times \text{len}} \in [0.8,\; 1.2]\right)
\end{equation}

When detected, the estimate should be treated as a lower bound rather than a point estimate: the true NDV is at least as high as the dictionary threshold, but the inversion equation no longer constrains it.

\section{Min/Max Diversity Estimation}
\label{sec:minmax}

\subsection{Row Group Extrema as Implicit Samples}

Each row group computes and stores the minimum and maximum values for each column. Across $n$ row groups, let $m_{\min}$ denote the number of distinct min values and $m_{\max}$ the number of distinct max values. These counts reflect how many different values appear as row group extrema--a signal that depends on the underlying cardinality and the data distribution.

\subsection{Coupon Collector Model}

The coupon collector problem models sampling with replacement from a finite population~\cite{feller}. Drawing $k$ items uniformly from $N$ distinct values, the expected number of distinct values observed is:
\begin{equation}
E[\text{distinct}] = N \times (1 - e^{-k/N})
\label{eq:coupon}
\end{equation}

We model the $n$ row group minima as $n$ draws from the population of $\text{NDV}$ distinct values:
\begin{equation}
E[m_{\min}] = \text{NDV} \times (1 - e^{-n/\text{NDV}})
\end{equation}

An analogous equation holds for $m_{\max}$.

\subsection{Inverting for NDV}

Given an observed count $m$ (either $m_{\min}$ or $m_{\max}$), we solve:
\begin{equation}
g(\text{NDV}) = \text{NDV} \times (1 - e^{-n/\text{NDV}}) - m = 0
\end{equation}

via Newton-Raphson with derivative:
\begin{equation}
g'(\text{NDV}) = 1 - e^{-n/\text{NDV}} \cdot \left(1 + \frac{n}{\text{NDV}}\right)
\end{equation}

We compute separate estimates from $m_{\min}$ and $m_{\max}$ and retain the larger of the two.

\subsection{Accuracy on Sorted Data}

For uniformly distributed data, row group minima cluster near the global minimum--we are sampling order statistics, not sampling uniformly from the value space. This causes min/max diversity to underestimate NDV.

For sorted or partitioned columns, each row group covers a distinct value range. The minima and maxima are spread across the value space, producing more distinct extrema and tighter estimates:

\begin{verbatim}
Uniform column (ndv=100, 50 row groups):
  Many row groups share similar min/max
  distinct_mins ~ 40

Sorted column (ndv=100, 50 row groups):
  Each row group covers a different range
  distinct_mins ~ 50
\end{verbatim}

Min/max diversity produces better estimates precisely where dictionary inversion fails.

\section{Distribution Detection}
\label{sec:distribution}

To select the appropriate estimator, we classify each column's data layout from the row group range patterns $(\min_i, \max_i)_{i=0}^{n-1}$.

\subsection{Metrics}

\paragraph{Range overlap.} We measure how much consecutive row group ranges intersect:
\begin{align}
\text{overlap}(r_i, r_{i+1}) &= \max(0,\; \min(\max_i, \max_{i+1}) - \max(\min_i, \min_{i+1})) \\
\text{overlap\_ratio} &= \sum_i \text{overlap}(r_i, r_{i+1}) \;/\; \text{total\_span}
\end{align}

\paragraph{Monotonicity.} We track midpoint progression across row groups:
\begin{equation}
\text{midpoint}_i = \frac{\min_i + \max_i}{2}, \qquad \text{monotonicity} = 1 - \frac{\text{sign\_changes}(\Delta\text{midpoints})}{n - 2}
\end{equation}

\subsection{Classification}

\begin{itemize}
\item \textbf{Sorted}: non-overlapping ranges (overlap\_ratio $< 0.1$) and high monotonicity ($> 0.9$)
\item \textbf{Pseudo-sorted}: moderate overlap ($< 0.3$) with clear drift (monotonicity $> 0.7$)
\item \textbf{Well-spread}: heavy overlap ($> 0.7$)--most distinct values appear in most row groups; dictionary inversion is reliable
\item \textbf{Mixed}: otherwise--both estimates used conservatively
\end{itemize}

\section{Combining Estimates}
\label{sec:combining}

\subsection{Hybrid Formula}

Given the complementary failure modes summarized in Table~\ref{tab:profiles}, the final NDV estimate takes the maximum of both methods, bounded by the number of non-null values:
\begin{equation}
\text{ndv\_final} = \min\bigl(\max(\text{ndv\_dict},\; \text{ndv\_minmax}),\; N - \text{nulls}\bigr)
\end{equation}

Each method underestimates in different regimes, so the higher estimate is more likely correct.

\subsection{Type-Specific Bounds}

For integer and date types, the value range provides a deterministic upper bound:
\begin{equation}
\text{ndv} \leq \min(\max - \min + 1,\; N - \text{nulls})
\end{equation}

For single-byte strings (status codes, flags):
\begin{equation}
\text{ndv} \leq \min({\sim}128,\; N - \text{nulls}) \quad \text{(printable ASCII)}
\end{equation}

\subsection{Schema Constraints}

When catalog metadata is available, schema constraints can provide tighter bounds than estimation. A foreign key column satisfies $\text{ndv} \leq \text{row\_count}(\text{referenced table})$.

\section{Batch Memory Estimation}
\label{sec:batch}

GPU query engines process data in batches. Given a global NDV estimate, we predict the dictionary memory required for a batch of size $B$ bytes without reading the batch.

Applying the coupon collector model (Equation~\ref{eq:coupon}), a batch containing $B / \text{len}$ rows drawn from a dictionary of size $D_{\text{global}} = \text{ndv} \times \text{len}$ requires:
\begin{equation}
D_{\text{batch}} = D_{\text{global}} \times (1 - e^{-B / D_{\text{global}}})
\end{equation}

For $n_{\text{batches}} = (N - \text{nulls}) \times \text{len} \,/\, B$, the total memory across all batches is:
\begin{equation}
D_{\text{total}} = \frac{(N - \text{nulls}) \times \text{len}}{B} \times D_{\text{global}} \times (1 - e^{-B / D_{\text{global}}})
\end{equation}

\paragraph{Limitation.} This assumes well-spread data, where each batch sees a representative sample of distinct values. For sorted data, each batch contains a distinct value subset and may require a dictionary approaching $D_{\text{global}}$--the coupon collector model does not apply.

\section{Applicability to Other Formats}

The technique requires two metadata features: (1)~dictionary encoding with uncompressed size reporting, and (2)~partition-level min/max statistics.

\begin{table}[h]
\centering
\begin{tabular}{lccc}
\toprule
Format & Dictionary Size & Partition Stats & Applicable \\
\midrule
Parquet & Yes & Yes (row group) & Yes \\
ORC & Yes & Yes (stripe) & Yes \\
F3~\cite{f3} & Likely & Likely & Likely \\
\bottomrule
\end{tabular}
\end{table}

F3 decouples dictionary scope from row groups (IOUnits), offering finer-grained control than Parquet. Statistics support is expected but not yet fully documented~\cite{f3}.

\section{Evaluation}

\subsection{Production Deployment}

The technique was implemented and deployed in Theseus~\cite{theseus}, a distributed GPU-accelerated query engine. NDV estimates drove a memory-based cost model for aggregate pushdown decisions, enabling partial aggregates to be pushed below joins to reduce data volume before GPU operations.

Production results on real-world Parquet datasets showed high accuracy, with errors typically below 10\% for well-spread columns. Sorted columns exhibited systematic underestimation via dictionary inversion alone, which the min/max diversity estimator addressed effectively. The hybrid approach proved robust across the distribution types encountered in practice.

The implementation and detailed experimental data were lost when VoltronData's assets were liquidated. Reproduction on public benchmarks is planned.

\subsection{Complexity}

\begin{table}[h]
\centering
\begin{tabular}{lcc}
\toprule
Operation & Time & Space \\
\midrule
Metadata parsing & $O(n)$ & $O(1)$ \\
Dictionary inversion & $O(1)$ & $O(1)$ \\
Min/max diversity & $O(n)$ & $O(1)$ \\
Length estimation & $O(n)$ & $O(1)$ \\
Distribution detection & $O(n)$ & $O(1)$ \\
\bottomrule
\end{tabular}
\end{table}

\noindent Where $n$ is the number of row groups. All operations are single-pass over the metadata. Distinct min/max counting uses a cardinality sketch (HyperLogLog) for $O(1)$ space; length estimation uses a length histogram bounded by the number of distinct lengths observed.

\section{Related Work}

\noindent\textbf{Query optimization.} Using cardinality statistics for cost-based planning dates to System~R~\cite{selinger}, which introduced index cardinality (ICARD) for selectivity estimation. NDV remains central to modern optimizers.

\noindent\textbf{Cardinality estimation.} HyperLogLog~\cite{hll} provides near-optimal $O(1)$-space approximate counting, and CVM~\cite{cvm} offers a simpler streaming alternative. Both require processing the data stream. Sampling-based methods~\cite{haas} trade accuracy for reduced scans. Our approach extracts estimates from existing metadata without additional storage or data access.

\noindent\textbf{Columnar statistics.} Prior work uses row group statistics for predicate pushdown~\cite{cloudera} and evaluates columnar storage formats~\cite{zeng}. We repurpose the same statistics for cardinality estimation.

\noindent\textbf{Dictionary encoding.} Studies model encoding efficiency and compression trade-offs~\cite{nvidia} but do not consider inverting metadata for NDV estimation.

\section{Conclusion}

Columnar file metadata encodes sufficient information to estimate column cardinality without accessing data pages. By inverting the dictionary size equation and treating row group min/max values as implicit cardinality sketches, we achieve zero-cost NDV estimation. Distribution detection routes between the two complementary estimators, providing robustness across data layouts.

Production deployment in Theseus demonstrated that the technique yields estimates accurate enough to drive cost-based optimization in a GPU-accelerated distributed query engine, with errors typically below 10\% on the data distributions encountered in practice. The same approach applies to any columnar format with dictionary encoding and partition statistics.

\appendix
\section{Core Equations}

\textbf{Dictionary size inversion:}
\begin{equation}
S = \text{ndv} \times \text{len} + (N - \text{nulls}) \times \lceil\log_2(\text{ndv})\rceil \,/\, 8
\end{equation}

\textbf{Coupon collector:}
\begin{equation}
m = \text{NDV} \times (1 - e^{-n/\text{NDV}})
\end{equation}

\textbf{Batch dictionary size:}
\begin{equation}
D_{\text{batch}} = D_{\text{global}} \times (1 - e^{-B/D_{\text{global}}})
\end{equation}

\end{document}